\newlength{\absize}
\documentclass[12pt]{article}
\usepackage{latexsym}
\usepackage{amssymb}
\usepackage{setspace}
\newcommand{\tr}{\mathop{\rm Tr}\nolimits}
\input prepictex
\input pictex
\input postpictex
\newdimen\tdim
\tdim=\unitlength
\def\stpltsmbl{\setplotsymbol ({\small .})}

\newbox\sru
\setbox\sru=\hbox{\beginpicture
\setcoordinatesystem units <\tdim,\tdim>
\stpltsmbl
\setquadratic
\plot
  0.0   0.0
  4.8   1.5
  7.5   5.0
  7.3   8.5
  5.0  10.0
  2.7   8.5
  2.5   5.0
  5.2   1.5
 10.0   0.0
/
\endpicture}
\def\springru #1 #2 *#3 /{\multiput {\copy\sru}  at
#1 #2 *#3 10 0 /}

\newcommand{\be}{\begin{equation}}
\newcommand{\ee}{\end{equation}}
\newcommand{\bea}{\begin{eqnarray}}
\newcommand{\eea}{\end{eqnarray}}

\newcommand{\comment}[1]{}

\usepackage{epsf}

\usepackage{color}
\definecolor{blue}{rgb}{0,0,.7}
\definecolor{red}{rgb}{.7,0,0}
\definecolor{orange}{rgb}{1,.6,0}
\definecolor{purple}{rgb}{.4,0,.5}
\definecolor{brown}{rgb}{.4,.2,.1}
\definecolor{green}{rgb}{0,.3,0}
\definecolor{white}{rgb}{.9,.9,.9}
\definecolor{truewhite}{rgb}{1,1,1}
\definecolor{gray}{rgb}{.8,.8,.8}

\begin{document}

\thispagestyle{empty}
\pagestyle{empty}
\newcommand{\starttext}{\newpage\normalsize
 \pagestyle{plain}
 \setlength{\baselineskip}{3ex}\par
 \setcounter{footnote}{0}
 \renewcommand{\thefootnote}{\arabic{footnote}}
 }
\newcommand{\preprint}[1]{\begin{flushright}
 \setlength{\baselineskip}{3ex}#1\end{flushright}}
\renewcommand{\title}[1]{\begin{center}\LARGE
 #1\end{center}\par}
\renewcommand{\author}[1]{\vspace{2ex}{\large\begin{center}
 \setlength{\baselineskip}{3ex}#1\par\end{center}}}
\renewcommand{\thanks}[1]{\footnote{#1}}
\renewcommand{\abstract}[1]{\vspace{2ex}\normalsize\begin{center}
 \centerline{\bf Abstract}\par\vspace{2ex}\parbox{\absize}{#1
 \setlength{\baselineskip}{2.5ex}\par}
 \end{center}}

\title{Color Fields on the Light-Shell}
\author{
 Howard~Georgi,\thanks{\noindent \tt georgi@physics.harvard.edu}
Greg~Kestin,\thanks{\noindent \tt kestin@physics.harvard.edu} 
Aqil~Sajjad,\thanks{\noindent \tt sajjad@physics.harvard.edu} 
 \\ \medskip
Center for the Fundamental Laws of Nature\\
Jefferson Physical Laboratory \\
Harvard University \\
Cambridge, MA 02138
 }
\date{\today}
\abstract{We study the classical color radiation from very high energy
collisions that produce colored particles. In the extreme high energy
limit, the classical color fields are confined to a light-shell expanding at $c$
and are associated with a non-linear $\sigma$-model on the 2D light-shell
with specific symmetry breaking terms. We
argue that the 
quantum version of this picture exhibits asymptotic freedom and may be a
useful starting point for an effective light-shell
theory of the structure between the
jets at a very high energy collider.
}

\newpage
\starttext

Those of us who have had the pleasure of learning or teaching from Ed
Purcell's classic book on electricity and magnetism~\cite{Purcell} cannot
forget the evocative figure in chapter 5 illustrating how a pulse of
electromagnetic radiation emerges from a kink in the
field of a charge that starts and stops. In this note, we suggest that a
similar picture may yield a useful starting point for
a description of very high energy
collisions between hadrons. 

The idea is a simple one. At a collider, colorless incoming particles
(whether leptons or hadrons) interact in a very small space-time region and
colored constituents emerge at high energies in various directions. This is
quite analogous to a situation in classical electrodynamics in which high
speed charged particles emerge suddenly at a point from an initially neutral
distribution of charges. In classical electrodynamics, we know what happens
and how to calculate it. A ``light-shell'' of electromagnetic radiation is produced
at the collision event and expands at the speed of light.\footnote{The
light-shell is thus a
constant $t$ slice of the light cone of the initial space-time event.} Outside the
light-shell, there are no fields. Inside the light-shell the electric and magnetic 
fields of the produced
charged particles match continuously (though with Purcell's kink) onto the
$\vec E$ and $\vec B$
fields on the light-shell. These are ``transverse'' --- tangent to the
shell and perpendicular to its direction of  motion.

What we are interested in for the analogy to very high energy hadronic
collisions is the situation in which the produced charged particles have
very high energy and move essentially at the speed of light, thus keeping
up with the light-shell of radiation produced in the collision. We will
consider the extreme (and of course unrealistic) limit in which the collision occurs instantaneously and with infinite energy so the charged
particles move at the speed of light from an initial space-time point and
the light-shell is infinitly thin. 
In this limit, not only are there no electric and magnetic fields outside the
light-shell, but there are also none inside the light-shell. All of the physics resides
on the thin spherical light-shell expanding at the speed of light. 

We believe that a similar picture should apply for hadronic collisions at
very high energies, for a very short time after the collision.
In this case, the initial collision involves
hard QCD processes taking place at energies large compared to
the QCD scale. This produces
very high energy colored particles that fly apart at the
speed of light and these particles, along with the color electric and magnetic 
fields they produce will be confined to an expanding light-shell, just as in
the case of electromagnetism. We hope this picture may be useful to describe the
physics for the range of times between the very short time scale of the
initial collision and the ``long'' time scale of $1/\Lambda_{QCD}$.

In this paper, we flesh out this idea by
looking at classical color fields in the appropriate limit. We will argue
that the classical color electric fields on the light-shell can be related to a non-linear
$\sigma$-model on a static two dimensional sphere with the Goldstone
bosons playing the role of the potential field and with specific
symmetry breaking related to the color charges of the high energy particles
producing the fields. We
will further argue that the quantum mechanical description of these light-shell
fields likely exhibits
asymptotic freedom with a coupling $g(r)$ depending on the radius of the
light-shell, with~\cite{Polyakov:1975rr}
\begin{equation}
\frac{1}{g(r)^2}\propto\log\left(\frac{1}{r\Lambda_{QCD}}\right)
\end{equation}
for $r\ll1/\Lambda_{QCD}$. As the light-shell expands, the QCD interactions
become more and more important until we reach a radius of the order of the
QCD scale, at which point perturbation theory breaks down. We hope that this connection with the non-linear $\sigma$-model will be
another useful result of this work. Field theorists have long studied the
analogies between non-Abelian gauge theories in $3+1$ dimensions and
non-linear $\sigma$-models in $2$ dimension, making use of some the
powerful tools available in the smaller number of dimensions (see for
example, \cite{Wiegmann:1984pw}). We argue that this is 
not just an analogy. The non-linear $\sigma$-model {\bf IS} QCD in an appropriate
limit. We hope that eventually, this will allow some of the magic of 2D field theories to be brought to bear on the physics of jets in high energy collisions.

There have also been some interesting works in related directions. In~ \cite{Verlinde:1993te}, a simplified effective theory for QCD is derived in the high-energy limit. While this effective theory is still ($3+1$)-dimensional, its interactions are described, to leading order, in terms of a 2-dimensional $\sigma$-model on the \emph{transverse plane}. Another interesting paper is~\cite{McLerran:1993ni}, in which the classical equation for the gluon field is solved for the case in which the source is a delta function along the light-cone in the $z$ direction. This calculation has some resemblance with part of what we show in this note, except that we take the source to be a distribution of charges moving spherically outward from the origin along the $t=r$ light-shell instead of a delta function along a specific direction. Additionally, some of the recent work on assymptotic gauge symmetries has been exploring related themes involving the null sphere at infinity~\cite{He:2015zea, Pasterski:2015zua, Adamo:2014yya, Adamo:2015fwa, Campiglia:2015qka}.

In the case of classical E\&M, the light-shell picture can be verified directly by
solving Maxwell's equations. In this case the fields on the light-shell are
static free fields (and of course there is no asymptotic freedom). Specifically, consider charges $q_i$ satisfying $\sum_i q_i$ starting at the origin at $t=0$ and travelling at the speed of light in the $\hat n_i$ directions. It is possible to show that the potentials (after an appropriate gauge transformation) are given by
\begin{equation}
A^0\left(t,\vec r\,\right)=
\phi\left(t,\vec r\,\right)
= -\sum_i q_i \,\delta(t-r)\,\log\left(1-\hat n_i\cdot\hat r\right) 
\label{phi-retarded-lsg}
\end{equation}
and
\be
\vec A\left(t,\vec r\,\right)
= \hat r\,\phi\left(t,\vec r\,\right)
\label{A-retarded-lsg}
\ee
which are determined by the single function $\phi$. Note that these potentials satisfy the gauge condition
\be
v_\mu A^\mu = 0
\label{LSG-condition1}
\ee
where
\be
v^0=1\;\;\mbox{and}\;\; \vec v=\hat r
\label{vmu}
\ee
We call this the light-shell gauge (LSG) condition and it is an important part of our quantum effective field theory on the light-shell which we introduce in the simplified zero flavor setting of scalar QED in~\cite{LSETintro}. We give the calculation of the photon propagator in light-shell gauge in~\cite{LSGprop} and discuss radiative corrections which reproduce the familiar double log structure of the full theory in~\cite{LSETintro, LSETrunning}.

It is straightforward to calculate the electric and magnetic fields from the potentials (\ref{phi-retarded-lsg}) and (\ref{A-retarded-lsg}). We find that they are both parallel to the surface of the $t=r$ and are given by
\be
\vec E\left(t,\vec r\,\right)
=-\sum_i q_i\,\hat r\times\left(\hat r\times\hat n_i\right)\,\delta(t-r)\,
\frac{1}{r-\hat n_i\cdot\vec r}
\label{E-field-abelian}
\ee
\be
\vec B\left(t,\vec r\,\right)
=\sum_i q_i \,\hat r\times\hat n_i\,\delta(t-r)\,
\frac{1}{r-\hat n_i\cdot\vec r}
\label{B-field-abelian}
\ee

The non-Abelian case is more complicated, and it is not obvious how to write down and solve the relevant equations directly. Here we will adopt a less direct route by assuming a simple form for the gauge fields and imposing the physics of the collision. Specifically, we will start by assuming that the gauge fields are zero outside the $t= r$ sphere. We will then go on to construct the field strengths $\mathcal{F}^{\mu\nu}_{a}$, and impose the following two conditions:
\begin{enumerate} 
\item In the extreme relativistic limit, we expect no energy/momentum density inside the light-shell. Thus the field strengths must vanish for $r<t$, and lie entirely on the sphere.
\item The fields satisfy the non-abelian version of Maxwell's equations, which tell us how the charges on the light-shell produce the fields: 
\be
D_\nu\mathcal{F}^{\mu\nu}=4\pi\mathcal{J}^\mu
\label{maxwell}
\ee
where $\mathcal{J}^\mu$ is a color current density. 
\end{enumerate}

For implementing this plan, we are ultimately interested in color gauge fields of the form
\be
A_a^\mu(t,\vec r\,)=\xi^\mu_a(t,\vec r\,)\,\theta(t-r)
\label{discontinuous}
\ee
which drop to zero discontinuously at the light-shell. When we differentiate these gauge fields, we will find field strengths proportional to $\delta(t-r)$ --- that is to say confined to the light-shell. The basic idea is then to use (\ref{discontinuous}) to construct the field strengths and see what the dynamics of classical QCD tells us about the field strengths on the light-shell.

The form (\ref{discontinuous}) is simple and appealing, and in the Abelian case, it is actually good enough to reproduce the results of a direct calculation using retarded potentials. However, as we will see, to understand the non-Abelian equations of motion, it is important to think about getting to this singular situation as a limit of smoother gauge fields. We want to understand when and how our results depend on the details of how we go to the discontinuous limit. So we will think about obtaining (\ref{discontinuous}) as a limit of smooth gauge fields, $\mathcal{A}^\mu_{a}(\epsilon,t,\vec r\,)$, such that 
\be
\lim_{\epsilon\to0}\mathcal{A}^\mu_{a}(\epsilon,t,\vec r\,)
=\xi^\mu_a(t,\vec r\,)\,\theta(t-r)
\label{limit}
\ee

To construct the field strengths, we will need derivatives of this as well as products of more than one such field with different non-abelian group indices. For the derivatives, we will use
\be
\partial^\mu\theta(t-r)=v^\mu\delta(t-r)
\label{vmu2}
\ee
with $v^\mu$ defined in (\ref{vmu}). This gives the relation
\be
\lim_{\epsilon\to0}\partial^\nu\mathcal{A}^\mu_{a}(\epsilon,t,\vec r\,)
=\theta(t-r)\,\partial^\nu\xi^\mu_a(t,\vec r\,)
+\delta(t-r)\,v^\nu\,\xi^\mu_a(t,\vec r\,)
\label{partialA}
\ee
For a product of two such fields without derivatives, we can write 
\be
\lim_{\epsilon\to0}\mathcal{A}^\mu_{a}(\epsilon,t,\vec r\,)
\,\mathcal{A}^\nu_{b}(\epsilon,t,\vec r\,)
=\xi^\mu_a(t,\vec r\,)\,\xi^\nu_b(t,\vec r\,)\,\theta(t-r)
\label{AA}
\ee

In the field strength, the relations (\ref{partialA}) and (\ref{AA}) are all we need, and we find for $\epsilon\to0$
\be
\mathcal{F}^{\mu\nu}_{a}
=\partial^\mu\mathcal{A}^\nu_{a}
-\partial^\nu\mathcal{A}^\mu_{a}
+f_{abc}\mathcal{A}^\mu_{b}
\mathcal{A}^\nu_{c}
\ee
\be
\to
\delta(t-r)\,\left(v^\mu\xi_a^\nu-v^\nu\xi_a^\mu\right)
+\theta(t-r)\,\left(\partial^\mu\xi_a^\nu-\partial^\nu\xi_a^\mu
+f_{abc}\xi_b^\mu\xi_c^\nu\right)
\label{F}
\ee
Note that we have normalized the gauge fields to behave simply under the non-Abelian gauge invariance, so that under a gauge transformation ($T_a=\lambda_a/2$ where the $\lambda_a$ are the Gell-Mann matrices)
\be
\mathcal{A}^\mu=\mathcal{A}_a^\mu T_a
\to U \mathcal{A}^\mu U^\dagger-i U\partial^\mu U^\dagger
\ee

We now apply the 1st of the two conditions we listed above (namely that the field strengths vanish inside the sphere). This means that in equation (\ref{F}), the coefficient of the theta functions must be zero:
\be
\partial^\mu\xi_a^\nu-\partial^\nu\xi_a^\mu
+f_{abc}\xi_b^\mu\xi_c^\nu=0
\label{insidexi}
\ee
We then have field strengths only on the light-shell
\be
\mathcal{F}^{\mu\nu}_{a}
\to {F}^{\mu\nu}_{a}
=\delta(t-r)\,\left(v^\mu\xi_a^\nu-v^\nu\xi_a^\mu\right)
\label{F-only-on-sphere}
\ee

We now apply to this field strength the second condition, namely, that the fields satisfy the non-abelian Maxwell's equations) (\ref{maxwell}).  While doing so, we will also get some terms containing derivatives of delta functions, and will assume that these must vanish.
 
On the left hand side of (\ref{maxwell}), we encounter two interesting things. In color components, it
can be divided into four terms as follows.
\begin{equation}
\partial_\nu\left(\partial^\mu\mathcal{A}^\nu_{a}
-\partial^\nu\mathcal{A}^\mu_{a}\right)
+\partial_\nu\left(f_{abc}\mathcal{A}^\mu_{b}\mathcal{A}^\nu_{c}
\right)
+f_{ade}\mathcal{A}_{d\nu}
\left(\partial^\mu\mathcal{A}^\nu_{e}
-\partial^\nu\mathcal{A}^\mu_{e}\right)
+f_{ade}\mathcal{A}_{d\nu}\left(
f_{ebc}\mathcal{A}^\mu_{b}\mathcal{A}^\nu_{c}
\right)
\label{lhs}
\end{equation}
The $\epsilon\to0$ limits of the second and fourth terms in (\ref{lhs}) are
straightforward, respectively
\begin{equation}
\partial_\nu\left(\theta(t-r)\,f_{abc}\xi^\mu_{b}\xi^\nu_{c}\right)
\label{2nd}
\end{equation}
\begin{equation}
\theta(t-r)\,f_{ade}\xi_{d\nu}\left(
f_{ebc}\xi^\mu_{b}\xi^\nu_{c}\right)
\label{4th}
\end{equation}
The first term can be written as a sum of three terms:
\begin{equation}
\partial^0\left(\delta(t-r)\,v_\nu\,\left(v^\mu\xi_a^\nu-v^\nu\xi_a^\mu\right)\right)
\label{d0delta}
\end{equation}
\begin{equation}
+\delta(t-r)\,(v_0\partial_\nu-\partial_0v_\nu)\left(v^\mu\xi_a^\nu-v^\nu\xi_a^\mu\right)
\label{1delta}
\end{equation}
\begin{equation}
+\partial_\nu\left(
\theta(t-r)\,\left(\partial^\mu\xi_a^\nu-\partial^\nu\xi_a^\mu\right)
\right)
\label{1theta}
\end{equation}
The last of these, (\ref{1theta}), combines with (\ref{2nd}) to give zero by
virtue of (\ref{insidexi}).
The first must vanish if we are to avoid derivatives of
$\delta$-functions, which implies (because $v_\mu v^\mu=0$)
\begin{equation}
\delta(t-r)\,v_\mu\xi^\mu=0
\label{transversexi}
\end{equation}	
Comparing with (\ref{F-only-on-sphere}), you can see that this is the condition that the
color electric field on the light-shell is tangent to the light-shell, perpendicular to
the direction of motion of the light-shell, $\hat r$. We expected this
on physical grounds, and we now see that it is necessary for the
consistency of the picture. Comparing (\ref{transversexi}) with (\ref{limit}) also tells us that the gauge field in the limit $\lim_{\epsilon\to0}\mathcal{A}^\mu_{a}(\epsilon,t,\vec r\,)$ satisfies the light-shell gauge at least on the sphere.

Finally, we consider the third term in (\ref{lhs}). This term is
problematic because it is {\bf not determined} by the limiting value of
$\mathcal{A}^\mu$. The total derivative of a product of
$\mathcal{A}^\mu$s is determined,
\begin{equation}
\left(
\mathcal{A}^\nu_{b}(\epsilon,t,\vec r\,)
\,\partial^\lambda\mathcal{A}^\mu_{a}(\epsilon,t,\vec r\,)
+\mathcal{A}^\mu_{a}(\epsilon,t,\vec r\,)
\,\partial^\lambda\mathcal{A}^\nu_{b}(\epsilon,t,\vec r\,)
\right)
\to
\partial^\lambda\left(
\theta(t-r)\,\xi^\mu_a(t,\vec r\,)\xi^\nu_b(t,\vec r\,)
\right)
\label{bothterms}
\end{equation}
However, for the product of one $\mathcal{A}$ with the derivative of
another, the limit depends on the details of their shapes.
In general we can write
\begin{equation}
\begin{array}{c}
\mathcal{A}^\mu_{a}(\epsilon,t,\vec r\,)
\,\partial^\lambda\mathcal{A}^\nu_{b}(\epsilon,t,\vec r\,)\to
\\
\displaystyle
\theta(t-r)\,\xi^\mu_a(t,\vec r\,)\,\partial^\lambda\xi^\nu_b(t,\vec r\,)
+\delta(t-r)\left(\frac{1}{2}v^\lambda\,\xi^\mu_a(t,\vec r\,)\,\xi^\nu_b(t,\vec r\,)
+\kappa^{\mu\lambda\nu}_{ab}(t,\vec r\,)\right)
\end{array}
\label{extra}
\end{equation}
where
\begin{equation}
\kappa^{\mu\lambda\nu}_{ab}(t,\vec r\,)
=
-\kappa^{\nu\lambda\mu}_{ba}(t,\vec r\,)
\label{extra2}
\end{equation}
The $\kappa$ term is the most general thing we
can write down consistent with (\ref{bothterms}).\footnote{Note
that this ambiguity only appears in the non-Abelian theory because of the
non-linearity of the equations of motion. There is no $\kappa$ in E\&M.} 
Using (\ref{extra}), we
get for the third term in (\ref{lhs})
\begin{equation}
\theta(t-r)\,
f_{abc}\,\xi_{b\nu}\left(\partial^\mu\xi^\nu_{c}
-\partial^\nu\xi^\mu_{c}\right)
+\delta(t-r)\,\kappa_a^{\mu}
\label{3rd}
\end{equation}
where
\begin{equation}
\kappa_a^{\mu}=f_{abc}\,g_{\lambda\nu}
\left(\kappa^{\lambda\mu\nu}_{bc}-\kappa^{\lambda\nu\mu}_{bc}\right)
\end{equation}
and we have used (\ref{transversexi}) and the antisymmetry of $f_{abc}$ to set
\begin{equation}
\delta(t-r)\,\frac{1}{2}\,
f_{abc}\,\xi_{b\nu}\left(v^\mu\xi^\nu_{c}
-v^\nu\xi^\mu_{c}\right)
=0
\label{zero}
\end{equation}
We will see later that something crucial happened in (\ref{3rd}). The
explicit non-linear dependence on $\xi$ in (\ref{zero}) goes away, but the
$\kappa$ term remembers the non-linear form of the field equations.
We will argue later that this extra $\kappa$ term is necessary for the consistency
of the picture.
Putting all this together, again using (\ref{insidexi}), Maxwell's equations become
\begin{equation}
\delta(t-r)\,\Biggl[
\left(v_0\partial_\nu-v_\nu\partial_0\right)
\left(v^\mu\xi^\nu-v^\nu\xi^\mu\right)
+\kappa_a^{\mu}-4\pi\sigma_a\,v^\mu\Biggr]=0
\label{deltaterm}
\end{equation}

We are interested in what these equations tell us about the fields on the light-shell, so we will eliminate $t$ and evaluate (whenever we can) the fields
for $t=r$. Look for example at $\mu=0$ in (\ref{deltaterm}).
\begin{equation}
\delta(t-r)\,\Biggl[
\left(
\vec\nabla+\hat r\partial_0
\right)\cdot
\left(\vec\xi_a(t,\vec r\,)-\hat r\xi_a^0(t,\vec r\,)\right)
+\kappa_a^0-4\pi\sigma_a\Biggr]=0
\label{mu0}
\end{equation}
Define ``light-shell fields'' which are functions only of $\vec r$ by setting
$t=r$ to go onto the light-shell:
\begin{equation}
\vec e_a(\vec r\,)\equiv \left.\left(\vec\xi_a(t,\vec r\,)-\hat
r\,\xi_a^0(t,\vec r\,)\right)\right|_{t=r} 
\label{e}
\end{equation}
Then because of (\ref{transversexi}), these fields are transverse,
\begin{equation}
\hat r\cdot\vec e_a(\vec r\,)=0
\label{transverse}
\end{equation}
In terms of $\vec e$, (\ref{mu0}) becomes
\begin{equation}
\delta(t-r)\,\left(
\vec\nabla\cdot\vec e_a(\vec r\,)-4\pi\sigma_a(\vec r\,)+\kappa_a^0(\vec
r\,)
\right)=0
\label{mu0shell}
\end{equation}
Notice that the derivatives of $\xi$ with respect to $\vec r$ and $t$ have
conspired to give derivatives of the light-shell fields just with respect to
$\vec r$. Because (\ref{mu0shell}) is true for all $t$, we must have
\begin{equation}
\vec\nabla\cdot\vec e_a(\vec r\,)=4\pi\sigma_a(\vec r\,)-\kappa_a^0(\vec r\,)
\label{electric}
\end{equation}
Thus $\vec e_a$ is a kind of electric field on the light-shell, but
(\ref{electric}) is true in a static 3D space.\footnote{You might wonder what becomes
of the color gauge invariance, since it looks like the gauge field $\xi$ is
simply turning into the gauge invariant field strength, $\vec e_a$. The
answer is that gauge transformations that preserve the form
(\protect\ref{discontinuous}) of $A^\mu$ change the $\xi$s inside but do
not change the light-shell fields, $\vec e_a$ except for global color
rotations, which of course remain.}   

For the space components of (\ref{deltaterm}), a similar manipulation gives
\begin{equation}
\vec\nabla\times\left({\hat r}\times\vec e_a\,\right)
=4\pi\sigma_a\,{\hat r}-\vec\kappa_a
\label{magnetic}
\end{equation}
This is very reasonable. It says that the curl of the magnetic field on the
light-shell is related to the current and $\vec\kappa$.
We can combine (\ref{magnetic}) and  (\ref{electric}), to obtain
\begin{equation}
{\hat r}\times\left(\vec\nabla\times\vec e_a\,\right)
=\left(\hat r\,\kappa_a^0-\vec\kappa_a\right)
\label{kappaconstraint}
\end{equation}
We will see shortly that this gives a constraint on $\vec\kappa$.

In electromagnetism, in spite of the singularity of (\ref{F-only-on-sphere}),
we can give direct physical meaning to the light-shell fields.
$\vec e$ is the impulse per unit
charge produced by the light-shell as it passes by a stationary 
infinitesimal test charge. This is
finite and 
independent of the detailed shape of the field as the shell width goes to
zero. It is not so
obvious that this concept makes sense in the non-Abelian case, because we
cannot make an arbitrarily small test charge. It appears that
to construct gauge invariant quantities that are finite in the
$\epsilon\to0$ limit, we have to take ratios. For example the surface
energy density on the light shell goes to $\infty$ as $\epsilon\to0$, but
ratios of energy densities at different points should be finite. 

Now let's look in more detail at the vanishing of the field in the interior
and see what part of this we can write in terms of light-shell fields. We know from the vanishing of the field for $r<t$ that
\begin{equation}
\nabla^j\xi^k_a-\nabla^k\xi^j_a=f_{abc}\xi^j_b\xi^k_c
\quad\mbox{and}\quad
\partial^0\xi^j_a
+\nabla^j\xi^0_a=-f_{abc}\xi_b^0\xi^j_c
\end{equation}
We can combine these into light-shell fields as follows:
\begin{equation}
(\nabla^j+{\hat r}^j\partial^0)\left(\xi_a^k-{\hat r}^k\,\xi_a^0\right)
-(\nabla^k+{\hat r}^k\partial^0)\left(\xi_a^j-{\hat r}^j\,\xi_a^0\right)
=f_{abc}\left(\xi_b^j-{\hat r}^j\,\xi_b^0\right)\times
\left(\xi_c^k-{\hat r}^k\,\xi_c^0\right)
\end{equation}
so we can use (\ref{e}) and 
set $t=r$ and conclude that the 3D theory of $\vec e_a$ satisfies
\begin{equation}
\nabla^j e^k_a-\nabla^k e^j_a=f_{abc} e^j_b e^k_c
\quad\mbox{or}\quad
\vec\nabla\times\vec e_a=\frac{1}{2}f_{abc}\,\vec e_b\times\vec e_c
\label{inside}
\end{equation}

(\ref{inside}), all by itself, has a number of consequences. Because the
$\vec e_a$s are perpendicular to $\hat r$, their cross product must be in
the $\hat r$ direction. Thus 
\begin{equation}
\hat r\times\left(\vec\nabla\times\vec e_a\right)
=0
\label{radialcrossproduct1}
\end{equation}
But if we take the gradient of (\ref{transverse}) and simplify, we get 
\begin{equation}
\hat r\times\left(\vec\nabla\times\vec e_a\right)
= -\frac{1}{r}\left(1+\vec r\cdot\vec\nabla\,\right)\vec e_a
\label{radialcrossproduct2}
\end{equation}
And on comparing this with (\ref{radialcrossproduct1}), we see that $\vec e_a$ scales trivially,
\begin{equation}
\left(\vec  r\cdot\vec \nabla\right)\vec e_a=-\vec e_a
\label{trivialscaling}
\end{equation}
Thus $\vec e_a$ is just $1/r$ times a vector function of $\hat r$.
Again, this follows directly from (\ref{inside}) which in turn follows from
the vanishing of the fields inside the light-shell.
(\ref{radialcrossproduct1}) together with (\ref{kappaconstraint}) also implies
\begin{equation}
\kappa_a^\mu=v^\mu\kappa_a
\label{kappav}
\end{equation}
for some scalar function $\kappa_a$,
so that like the current, $\kappa_a^\mu\propto v^\mu$. 
Thus in the limit, all the information from the non-Abelian Maxwell's
equations is contained in (\ref{kappav}) and the following relations:
\begin{equation}
\vec\nabla\times\vec e_a=\frac{1}{2}f_{abc}\,\vec e_b\times\vec e_c
\quad\quad\quad
\hat r\cdot\vec e_a=0
\quad\quad\quad
\left(\vec  r\cdot\vec \nabla\right)\vec e_a=-\vec e_a
\label{together1}
\end{equation}
\begin{equation}
\vec\nabla\cdot\vec e_a=4\pi\sigma_a-\kappa_a\equiv4\pi\tilde\sigma_a
\label{together2}
\end{equation}
Notice that the effective charge density $4\pi\tilde\sigma_a$ must scale like $1/r^2$ (consistent with charge conservation).

We can solve (\ref{together1}) for the $\vec e_a$ fields as follows:
\begin{equation}
\vec e_aT_a=-i\,U(\hat r)^\dagger\vec\nabla U(\hat r)
\label{E-solution-paper1}
\end{equation}
where $U^\dagger U=I$ is a special unitary matrix.
Now trivial scaling and transversality 
are automatic because $U$ depends only on $\hat r$. (As an aside, (\ref{together2}) and (\ref{E-solution-paper1}) closely resemble equations 11 and 16, respectively, in \cite{McLerran:1993ni}.)

Because of (\ref{trivialscaling}), our picture is classically scale
invariant and
we could write the classical theory as
a purely two dimensional theory on the light-shell, and simply choose
$r=1$. Physically, however, it is sometimes convenient to think about the theory
as we actually use it, in the full three
dimensional space, but with the fields living on an expanding light-shell of radius $r=t$.

Having dealt with (\ref{together1}), we now want to find a Lagrangian that gives (\ref{together2}) as the equation of motion so that we can eventually do quantum mechanics. We have now eliminated time and are in purely Euclidean space, so this is just the energy. We would expect a contribution proportional to $\tr(\vec e\,^{2})$, which in
terms of $U$ can be written as (where $B$ is some geometrical constant that
we do not know how to calculate at this point, and $g$ is the dimensionless
coupling constant)
\begin{equation}
\frac{B}{g^2}\tr\left(\vec\nabla U(\hat r)^\dagger\cdot\vec\nabla U(\hat r)\right)
\label{l1}
\end{equation}
This is the Lagrangian for a
non-linear $\sigma$-model on the light-shell and the $U$ fields (which in some
sense are the potentials associated with the electric fields) are Goldstone
boson fields associated with the breaking of an $SU(3)_L\times SU(3)_R$,
$U\to LUR^\dagger$ down to the diagonal $SU(3)$, $U\to VUV^\dagger$. The
electric fields $\vec e_a$ are Noether currents associated with the $SU(3)_R$
symmetry, so if (\ref{l1}) were the whole story, $\vec e_a$ would be conserved, in
agreement with (\ref{together2}) without sources, for $\tilde\sigma=0$.
This is a renormalizable theory in 2D, and Polyakov showed long ago that the
coupling $g$ exhibits asymptotic freedom~\cite{Polyakov:1975rr}. What happens in this
particular geometrical situation is simple and
interesting. Because the fields live on the light-shell of radius $r$, the
momenta in the theory are actually angular momenta divided by $r$. The
$\ell=0$ mode is absent because it gives no contribution to $\vec e$ if the total net charge on the light-shell is zero. 
The momenta are
bounded away from zero and quantized in units of $1/r$. The infrared
divergence that one would expect in a flat 2D theory is cut off at
$r$. Because all the momenta scale with $1/r$, it is appropriate to
choose the renormalization scale to scale with $1/r$, so
the coupling depends on the radius.


Up to this point, we believe that our analysis is quite robust. In the
appropriate limit, we can describe the physics in terms of light-shell
fields, and the condition that the field strengths vanish inside the
light shell implies quite directly that these fields are described by a
non-linear $\sigma$-model. We are on shakier ground from here on, where we
discuss the dependence on the charges and currents of the high energy
particles that are producing the fields. Here $\kappa$ gets involved, and
in our indirect approach to the limit, we do not know exactly what $\kappa$
is. But we believe that a non-zero $\kappa$ is necessary and have a guess for
its form, and we will now discuss the reasons for the belief and the
guess.
Suppose first that $\kappa=0$. Then the equation of motion for $U$ would be
(from (\ref{together2})), 
\begin{equation}
\vec\nabla\cdot\left(-i\,U^\dagger\vec\nabla U\right)
=4\pi\sigma
\label{kappa0}
\end{equation}
where the right hand side is independent of $U$.
However, it is not possible to add to the Lagrangian (\ref{l1}) a term
$F(U)$ that gives this equation of motion, because Noether's theorem
requires that to get (\ref{kappa0}) from an infinitesimal symmetry transformation,
\begin{equation}
\delta U=U\,i\delta\zeta
\label{delta}
\end{equation}
we need
\begin{equation}
\delta F=4\pi\tr(\sigma\delta\zeta)
\label{want}
\end{equation}
To see why this is a problem, write $U$ in terms of unconstrained octet
components, 
$U=e^{i\Pi_aT_a}$ so (\ref{delta}) is
\begin{equation}
\delta\zeta=O_a\delta\Pi_a\quad\mbox{where}\quad O_a
\equiv -iU^\dagger\frac{\delta U}{\delta\Pi_a}
\end{equation}
Thus we want
\begin{equation}
\frac{\delta F}{\delta\Pi_a}=\frac{B}{g^2}4\pi \sigma
\,O_a
\label{want2}
\end{equation}
But 
\begin{equation}
\frac{\delta O_a}{\delta\Pi_b}-\frac{\delta O_b}{\delta\Pi_a}
=-i\biggl[O_a\,,\,O_b\biggr]\neq0
\end{equation}
which means that (\ref{want2}) is not consistent. In the presence of
$\kappa$, there are additional terms in $\delta F$ coming from the
dependence of $\kappa$ (and thus $\tilde\sigma$) on the $\Pi$s. One simple
possibility is 
\begin{equation}
4\pi\tilde\sigma
=4\pi\sigma-\kappa
=2\pi(\sigma U+U^\dagger\sigma)
\end{equation}
which would emerge in the equation of motion from the Lagrangian
\begin{equation}
\frac{B}{g^2}\tr\left(\vec\nabla U^\dagger\cdot\vec\nabla U
-2\pi i(\sigma U-U^\dagger\sigma)\right)
\label{l2}
\end{equation}
This our guess for the structure of the effective theory on the light-shell.

We believe that this analysis makes a very plausible case that very high energy
collisions involving colored particles can be described by a light-shell
effective field theory in which the dynamical fields are the Goldstone
bosons of a non-linear $\sigma$-model on the light-shell at $t=r$. 
To go further, we must go beyond our
indirect arguments and see how to construct the light-shell effective theory
directly from the underlying QCD theory. Then we should be able to 
do the perturbative matching onto the light-shell effective theory from the
QCD physics of the original high-energy collision and better understand the
physical meaning of our light-shell fields. Efforts in this direction are continuing. 

\section*{Acknowledgements}

We have benefited from discussions with Matthew Schwartz and
David Simmons-Duffin.

\bibliography{shell}

\end{document}